\documentclass[12pt]{iopart}
\usepackage{graphicx}
\usepackage{iopams}

\begin{document}

\title [Broken discrete and continuous symmetries in 2D spiral antiferromagnets]{Broken discrete and continuous symmetries in two dimensional spiral antiferromagnets}

\author{A Mezio, C N Sposetti, L O Manuel and A E Trumper}
\address {Instituto de F\'{\i}sica Rosario (CONICET) and
Universidad Nacional de Rosario, Boulevard 27 de Febrero 210 bis, (2000) Rosario, Argentina.}
\ead{trumper@ifir-conicet.gov.ar}

\begin{abstract}
We study the occurrence of symmetry breakings, at zero and finite temperatures, in the $J_1-J_3$ antiferromagnetic Heisenberg model on the square lattice  using Schwinger boson mean field theory.
For spin-$\frac{1}{2}$ the ground state breaks always the $SU(2)$ symmetry with a continuous quasi-critical transition at $J_3/J_1\sim0.38$, from  N\'eel to  spiral long range order, although local spin fluctuations considerations suggest an intermediate disordered regime around $0.35 \lesssim J_3/J_1 \lesssim 0.5$, in qualitative agreement with recent numerical results. At low temperatures we find  a $Z_2$ broken symmetry region  with short range spiral order characterized by an Ising-like nematic order parameter that compares  qualitatively  well with classical Monte Carlo results. At intermediate temperatures the phase diagram shows regions with  collinear short range orders: for $J_3/J_1<1$  N\'eel  $(\pi,\pi)$ correlations  and  for $J_3/J_1>1$ a novel phase consisting of four decoupled third neighbour sublattices with  N\'eel  $(\pi,\pi)$ correlations in each one. We conclude that the effect of quantum and thermal fluctuations is to favour collinear correlations even in the strongly frustrated regime.     
\end{abstract}

\submitto{\JPCM}
\maketitle

\section{Introduction}

The study of unconventional phases represents a central topic of strongly correlated electron systems. In frustrated quantum antiferromagnets (AF) the  interest is mainly focused on the  possible stabilization of two dimensional (2D) quantum spin liquids \cite{anderson87,misguich05,balents10} that preserve all the microscopic symmetries of the Hamiltonian. In fact, in the last years, there have been a great interest in the classification of different types of quantum spin liquids based on the projective symmetry group \cite{wen02,vishwanath06,messio13}. However, the concrete detection of such spin liquids  on realistic quantum spin models seems to be still a delicate issue \cite{balents10,powell11,balents12,sorella12,sorella13}. The source of this classification are the mean field wave functions based on the bosonic and fermionic representations for the spin operator, originally used in the context of large $N$ theories \cite{arovas88,auerbach94}. The bosonic representation (Schwinger bosons) has the advantage of describing magnetically ordered states \cite{sarker89,chandra90a, read91} --which are known in several cases-- while quantum spin liquids states can be described by both, bosonic and fermionic representations \cite{sorella12,sorella13,read91}.\\
\indent Another route for the search of unconventional phases due to magnetic frustration has been the study of finite temperature transitions involving the rupture of non-trivial discrete degrees of freedom. This kind of transitions have been extensively investigated in the context of  the frustrated $J_1-J_2$ Heisenberg model  \cite{misguich05,chandra88,richter10}. Here the magnetic phase breaks the discrete lattice rotation symmetry from $Z_4$ to $Z_2$ with an associated Ising variable
that gives a measure of the $(0,\pi)$ and $(\pi,0)$ magnetic correlations \cite{chandra90b} while rotational symmetry is unbroken, as dictated by Mermin-Wagner theorem \cite{mermin66}. Several analytical \cite{chandra90b,flint08} and numerical \cite{weber03} studies in the $J_1-J_2$ model have confirmed the occurrence of  a finite temperature transition to a $Z_2$
broken symmetry phase that belongs to the Ising universality class. Less explored, instead, has been the occurrence of such transition in the $J_1-J_3$ model where, in contrast to the original case, the spin correlations are of spiral type \cite{read91,locher90,ferrer93}. Classically, for $\small{J_3/J_1> 1/4}$, there are two degenerate incommensurate spiral ground states, ${\bf Q}=(Q,Q)$ and  $(Q,-Q)$, that are connected by a global rotation followed by a reflexion about $y$. Then, the global symmetry of the classical ground state 
is $O(3)\times Z_2$. Classical Monte Carlo calculations \cite{capriotti04} predicts that a $Z_2$ broken symmetry phase described by an Ising nematic order parameter (see below)  survives within the finite temperatures range (see inset of figure \ref{tdiagram}), being the transition also of the Ising universality class. 
On the other hand, at zero temperature, numerical studies for $S=\frac{1}{2}$ predict the existence of an intermediate disordered regime in the range $0.4 \lesssim J_3/J_1\lesssim 0.8$ with, probably, short range order (SRO) plaquette and  spiral regimes between long range (LRO) N\'eel and spiral phases \cite{leung96,mambrini06,reuther11}; while for the special case $J_3/J_1\simeq 0.5$ there is evidence of an homogeneous spin liquid state \cite{capriotti04a}.\\
\indent In order to complement the classical Monte Carlo results and to make contact with the zero temperature
quantum regime, it is important to investigate the interplay between quantum and thermal fluctuation at low temperatures within a confiable theory. In this sense,  it has been shown that the Schwinger boson mean field (SBMF) approach based on the two singlet bond operators scheme \cite{flint08,ceccatto93,trumper97,manuel98} works very well for several  frustrated models. In particular, for the triangular AF, we have recently shown that  the zero temperature energy spectrum \cite{mezio11} and the low temperature thermodynamics properties \cite{mezio12} predicted by numerical methods are correctly reproduced. In addition this mean field scheme provides a qualitative good description of the finite temperature Ising transition in the $J_1-J_2$ model \cite{flint08}.\\  
\indent Motivated by these results, in the present article, we investigate the occurrence of both, the zero temperature $SU(2)$ broken symmetry ground state and the  finite temperature $Z_2$ broken symmetry transition in the frustrated $J_1-J_3$ Heisenberg model, using the Schwinger boson mean filed theory.
For the zero temperature quantum phase diagram (figure \ref{fig1}) we show that the two singlet scheme of the SBMF  takes correctly into account the effect of frustration $J_3/J_1$  within the collinear phases leading to qualitative and  quantitative differences with respect to previous calculations based on a one singlet scheme \cite{read91}. Although for $S=\frac{1}{2}$ the $SU(2)$ symmetry is always broken with a continuous quasi-critical transition at $J_3/J_1\sim0.38$, from N\'eel to spiral long range order (figure \ref{fig2}), local spin fluctuations considerations allow us to estimate a disordered regime $0.35 \lesssim J_3/J_1 \lesssim0.5$ between N\'eel and spiral states in qualitative agreement with recent numerical results \cite{reuther11}.   
As soon as temperature increases the finite temperature phase diagram (figure \ref{tdiagram}) shows a $Z_2$ broken symmetry phase  characterized by finite Ising nematic order with the rotational invariance restored. The behavior of the critical temperature $T_c$ with frustration, signalled by the vanishing of the nematic order parameter, compares quite well with classical Monte Carlo predictions \cite{capriotti04}. 
As temperature is further increased two different temperature effects --before reaching the paramagnetic phase-- are observed: for $J_3/J_1< 1$ short range  N\'eel $(\pi,\pi)$ correlations are favored while for $J_3/J_1> 1$ there is an intermediate novel phase --we have named it $(\pi,\pi)_4$-- characterized by four  decoupled third neighbour sublattices with AF short range correlations each one.

\section{The Schwinger boson approach within the two singlet scheme }

The AF Heisenberg model on the square lattice with first $J_1$ and third $J_3$ neighbours interaction is defined as
\begin{equation}
\hat{H}= J_{1}\sum_{<ij>}\hat{{\bf S}}_i \cdotp \hat{{\bf S}}_j +J_{3}\sum_{<ik>}\hat{{\bf S}}_i \cdotp \hat{{\bf S}}_k,
\label{j1j3model}
\end{equation}

\noindent where $<ij>$ and $<ik>$ denotes first and third neighbours, respectively, on the square lattice. In using the Schwinger boson representation 
for the spin operators \cite{auerbach94},

\begin{equation}
\hat{{\bf S}}_i= \frac{1}{2}\;{\bf b}^{\dagger}_i \; \vec{\sigma}\; {\bf b}_i,
\label{representation}
\end{equation}

\noindent with ${\bf b}^{\dagger}_i =({b}^{\dagger}_{i\uparrow}; {b}^{\dagger}_{i\downarrow})$ a spinor composed by the bosonic spin-$\frac{1}{2}$ operators ${b}^{\dagger}_{i\uparrow}$ and ${b}^{\dagger}_{i\downarrow}$ and
$\vec{\sigma}=(\sigma^x,\sigma^y,\sigma^z)$ the Pauli matrices, the condition of $2S$ boson per site

\begin{equation}
{b}^{\dagger}_{i\uparrow} {b}_{i\uparrow}+{b}^{\dagger}_{i\downarrow} {b}_{i\downarrow}=2S
\label{constraint}
\end{equation}

\noindent must be satisfied  in order to guarantee the physical Hilbert space. After replacing (\ref{representation}) in the spin-spin interaction terms  of (\ref{j1j3model}) they can be written in the following  two singlet bond operator scheme  

\begin{equation}
\hat{{\bf S}}_i \cdot \hat{{\bf S}}_j=: \hat{B}^{\dagger}_{ij} \hat{B}_{ij}:  -\hat{A}^{\dagger}_{ij}\hat{A}_{ij},
\label{int}
\end{equation}

\noindent with $i$ and $j$ representing either first or third neighbour sites, and the singlet bond operators $\hat{A}_{ij}$ and $\hat{B}_{ij}$ are defined as 

\begin{equation}
\hat{A}^{\dagger}_{ij}=\frac{1}{2}\sum_{\sigma}\sigma {b}^{\dagger}_{i \sigma}
{b}^{\dagger}_{j \bar{\sigma}},\;\;\;\;\hat{B}^{\dagger}_{ij}=\frac{1}{2}\sum_{\sigma}{b}^{\dagger}_{i\sigma}{b}_{j \sigma}.
\label{singlets}
\end{equation}

\noindent We call them singlets because they are rotationally invariant under $SU(2)$ transformations of the spinor ${\bf b}^{\dagger}_i =({b}^{\dagger}_{i\uparrow}; {b}^{\dagger}_{i\downarrow})$. The biquadratic terms of (\ref{int}) are related to the spin operators as

\begin{eqnarray}
 \hat{A}^{\dagger}_{ij}\hat{A}_{ij}&=& \frac{1}{4}(\hat{\bf S}_i-\hat{\bf S}_j)^2-\frac{S}{2} \\
 :\hat{B}^{\dagger}_{ij}\hat{B}_{ij}:&=&\frac{1}{4}(\hat{\bf S}_i+\hat{\bf S}_j)^2 -\frac{S}{2}\nonumber.
\end{eqnarray}
  
\noindent Then, after a mean field decoupling of the above expressions, the mean value of the operators $\hat{A}^{\dagger}_{ij}$ and $\hat{B}^{\dagger}_{ij}$  can be immediately associated to antiferromagnetic and ferromagnetic correlations between sites $i$ and $j$, respectively. 
Using the identity $:\hat{B}^{\dagger}_{ij}\hat{B}_{ij}:+\hat{A}^{\dagger}_{ij}\hat{A}_{ij}=S^2$ it is possible to write down the spin interaction (\ref{int}) in terms of either singlet operators, $\hat{B}_{ij}$ or $\hat{A}_{ij}$, and study independently pure ferromagnetic or antiferromagnetic phases, respectively \cite{arovas88}. For frustrated systems, where quantum disordered phases are expected, there are two schemes of calculation: one takes advantage of the above identity and uses only $\hat{A}_{
ij}$ operators \cite{read91} while the other one keeps both, $\hat{B}_{ij}$  and $\hat{A}_{ij}$ operators \cite{ceccatto93}. In principle both schemes are equivalent but at the mean field level the two singlet bond scheme has shown to be quite more accurate to describe the magnetically ordered regions of several frustrated models \cite{ceccatto93,trumper97,mezio11,mezio12}. More recently, this scheme has been used  to explore the possible existence of completely symmetric \cite{vishwanath06,li12} and weakly symmetric --chiral-- spin liquid states \cite{messio13} within the context of the projective symmetry group. Therefore, the two singlet scheme seems to be a more proper and versatile framework to investigate ordered and spin liquid phases in a unified way.    
  

\subsection{The mean field decoupling}
 
\noindent Performing the standard procedure \cite{ceccatto93}, the spin-spin interaction (\ref{int}) is replaced in the Hamiltonian (\ref{j1j3model}) along with the introduction of a Lagrange multiplier $\lambda$ so as to fulfill on average the constraint (\ref{constraint}). After a mean field decoupling, with $A_{ij}=\langle\hat{A}_{ij}\rangle=\langle\hat{A}^{\dagger}_{ij}\rangle$ and $B_{ij}=\langle\hat{B}_{ij}\rangle=\langle\hat{B}^{\dagger}_{ij}\rangle$, and Fourier transforming the Schwinger bosons to $k$-space 
the quadratic mean field Hamiltonian results 

\begin{eqnarray}
\hat{H}_{MF}&=& \sum_{\bf k} [ (\gamma^B_{\bf k}+\lambda)(b^{\dagger}_{{\bf k}\uparrow}b_{{\bf k}\uparrow} + 
b^{\dagger}_{-{\bf k}\downarrow}b_{-{\bf k}\downarrow})+ \label{cuadratic}\\
& & \;\;\;\;\;\;+i \gamma^A_{\bf k} b^{\dagger}_{{\bf k}\uparrow}
b^{\dagger}_{-{\bf k}\downarrow}- i \gamma^A_{\bf k} b_{{\bf k}\uparrow}
b_{-{\bf k}\downarrow}] -E_{MF}-2S\lambda N_s  \nonumber
\end{eqnarray}

\noindent where
$$
E_{MF}=\frac{N_s}{2} \sum_{\delta} J_{\delta} [B^2_{\delta}-A^2_{\delta}]
$$

\noindent and 
$$
\gamma^B_{\bf k}= \frac{1}{2} \sum_{\delta} J_{\delta} B_{\delta} \cos {\bf k}\cdot \delta, \;\;\;\;\;\;\;\;\;
\gamma^A_{\bf k}= \frac{1}{2} \sum_{\delta} J_{\delta} A_{\delta} \sin {\bf k}\cdot \delta,
$$

\noindent with the sums going over all the vectors $\delta$ connecting the first and the third neighbours, $N_s$ is number of sites and  where real mean field parameters satisfying the relations $B_{\delta}=B_{-\delta}$ and $A_{\delta}=-A_{-\delta}$ has been assumed. The mean field Hamiltonian (\ref{cuadratic}) can be diagonalized by applying a Bogoliubov transformation 

\begin{centering}
\begin{eqnarray}
b_{{\bf k}\uparrow}&=& u_{\bf k} \alpha_{{\bf k}\uparrow}-v_{\bf k} \alpha^{\dagger}_{-{\bf k}\downarrow} \nonumber\\
b_{{\bf k}\downarrow}&=& u_{\bf k} \alpha_{{\bf k}\downarrow}+v_{\bf k} \alpha^{\dagger}_{-{\bf k}\uparrow},
\label{Bogo}
\end{eqnarray}
\end{centering}
\noindent with $u_{\bf k}= [\frac{1}{2}(1+\frac{(\gamma^B_{\bf k}+\lambda)}{\omega_{\bf k}} )]^{\frac{1}{2}}$ and  $v_{\bf k}= i \ {\it sig}(\gamma^A_{\bf k})[\frac{1}{2}(-1+\frac{(\gamma^B_{\bf k}+\lambda)}{\omega_{\bf k}} )]^{\frac{1}{2}}$  the Bogoliubov coefficients, resulting
\begin{equation}
\hat{H}_{MF}= \sum_{\bf k} \omega_{\bf k} \left[ \alpha^{\dagger}_{{\bf k}\uparrow} \alpha_{{\bf k}\uparrow}+
 \alpha^{\dagger}_{-{\bf k}\downarrow} \alpha_{-{\bf k}\downarrow} \right] + E_{MF}
\end{equation}

\noindent with the same free spinon dispersion relation for the up and down flavours
\begin{equation}
  \omega_{\bf k}=\sqrt{(\gamma^B_{\bf k}+\lambda)^2- (\gamma^A_{\bf k})^2}.
\label{omega}
\end{equation}

\noindent The mean field free energy is given by 
\begin{equation}
F = E_{MF} + T \sum_{{\bf k} \sigma} \ln \left(1-e^{-\beta \omega_{{\bf k} \sigma}}\right),
\end{equation}
and the self-consistent equations for the mean field parameters, $A_{\delta}$, $B_{\delta}$ and $\lambda$ yield

\numparts
\begin{eqnarray}
A_{\delta}&=& \frac{1}{2N_s}\sum_{\bf k} \; \frac{\gamma^{\rm A}_{\bf k}}{\omega_{\bf k}} \; \;\left( 1 + 2 \, n_{{\bf k}} \right)\;\; \sin{\bf k} \cdot 
{\delta}, \label{selfa} \\
B_{\delta}&=& \frac{1}{2N_s}\sum_{\bf k} \; \frac{\gamma^{\rm B}_{\bf k}+\lambda}{\omega_{\bf k}}\;\; \left( 1 + 2 \, n_{{\bf k}} \right)\;\; 
\cos{\bf k}\cdot {\delta}, \label{selfb} \\  
S+\frac{1}{2}&=& \frac{1}{2N_s}\sum_{\bf k} \; \frac{\gamma^{\rm B}_{\bf k}+\lambda}{\omega_{\bf k}}\;\; \left( 1 + 2 \, n_{{\bf k}} \right), \label{selfc} 
\end{eqnarray}
\endnumparts
with ${n}_{{\bf k}}=(e^{\beta {\omega}_{\bf k}}-1)^{-1}$ the bosonic occupation number.  The rotationally invariant nature of the SBMFT allows to study magnetically disordered phases at finite temperatures in agreement with the Mermin-Wagner theorem \cite{mermin66}. This is manifested in the temperature dependent gapped spinon dispersion $\omega_{\bf k}$, once the self consistent equations (12) are solved, preventing the appearance of infrared divergences in the theory. Nonetheless, as temperature decreases the magnetic structure factor, $S({\bf k})=\sum_{\bf R}  e^{i {\bf k}. {\bf R}}\langle \hat{S}_0 \!\!\cdot\! \!\hat{S}_{\bf R}\rangle$, develops a maximum at ${\bf Q}=2{\bf k}_{min}$  with ${\bf k}_{min}$ the minimum of the relation dispersion $\omega_{\bf k}$ \cite{auerbach94}. For $T\rightarrow0$, the leading order of this maximum is related to the squared magnetization and $\omega_{{\bf k}_{min}}$ as $S({\bf Q})=\frac{1}{2N_s} \frac{(\gamma^{\rm B}_{{\bf k}_{min}}+\lambda)^2}{\omega^2_{{\bf k}_{min}}}= \frac {N_s}{2}m^2$. In the next section it is shown how the rupture of the $SU(2)$ symmetry is described in the zero temperature limit. 
  
\subsection{The treatment of $SU(2)$ broken symmetry in a spiral ground state}\label{sec}
The occurrence of the $SU(2)$ broken symmetry ground state at $T=0$ is related to the condensation of the Schwinger bosons \cite{sarker89,chandra90a}. To clarify this point it is instructive to focus on the ground state wave function 
of a finite size $N_s$ system. Even with semiclassical mean field solutions the ground state is magnetically disordered with a finite size gap dispersion that behaves as $\omega_{\pm\frac{\bf Q }
 {2}}\!\!\sim\!\!\frac{1}{N_s}$. The positiveness of $\omega_{\bf k}$ for all ${\bf k}$ guarantees the diagonalization of (\ref{cuadratic}), implying a zero spinons occupation number in the magnetic ground state. Using the requirement that $\alpha_{{\bf k} \sigma}|\texttt{gs}\rangle=0$, it can be easily shown that the ground state is a singlet with the following Jastrow form,

\begin{equation}
|\texttt{gs}\rangle\; = e^{\sum_{\bf k} f_{\bf k} b^{\dagger}_{{\bf k} \uparrow} b^{\dagger}_{-{\bf k} \downarrow}}
|0\rangle_b,
\label{WF}
\end{equation}

\noindent where $f_{\bf k}= -v_{\bf k}/u_{\bf k}$ and $|0\rangle_b$ is the vacuum of Schwinger bosons $b$. In the thermodynamic limit $\omega_{\pm\frac{\bf Q } {2}}\rightarrow 0$ and $f_{\pm\frac{\bf Q } {2}}\rightarrow 1$, meaning that the ground state develops an infinite accumulation of spin up and down bosons at 
${\bf k}=\pm\frac{\bf Q }{2}$. Then, the ground state can be splitted as

$$
 |\texttt{gs}\rangle\;=|{\phi}_c\rangle |\texttt{gs}\prime\rangle,     
$$

\noindent where $|{\phi}_c\rangle$ represents the condensed part, and

$$
 |\texttt{gs}\prime\rangle\;=e^{ \small{\sum_{\small{{\bf k}\neq\pm {\bf Q}/2}} f_{\bf k} 
b^{\dagger}_{{\bf k} \uparrow} b^{\dagger}_{-{\bf k} \downarrow}} }
|0\rangle_b
$$

\noindent is the non-condensed, or normal, part of the ground state \cite{chandra90a}. Given that the starting point (\ref{WF}) is a singlet, the appearance of the condensate must be related to the rupture of the $SU(2)$ symmetry. Physically, this can be thought by considering the hypothetic process of switching on a  modulated magnetic field $h$ with pitch ${\bf Q}$, then taking the thermodynamic limit $N_s\rightarrow \infty$, and finally making the limit $h\rightarrow 0$ \cite{sarker89,auerbach94}. For instance, a coherent state  

\begin{equation}
 |{\phi}_c\rangle= e^{{\small{\sqrt{\frac{N_s m}{2}}}} \left({b}^{\dagger}_{\frac{\bf Q}{2} \uparrow} + {b}^{\dagger}_{-\frac{\bf Q}{2} \uparrow} + 
i {b}^{\dagger}_{\frac{\bf Q}{2} \downarrow} -i{b}^{\dagger}_{-\frac{\bf Q}{2} \downarrow}\right)}|0\rangle_b\\
\label{condens}
\end{equation}

\noindent thus selected gives a quantum spiral state with magnetization $m$ and spiral pitch ${\bf Q}$ lying in the  $x-z$ plane. In fact, 
the mean value of the spin operator in this state yields
$$
\langle\texttt{gs}|\hat{S}_i^x|\texttt{gs}\rangle\\;= m \sin ({\bf Q}\cdot{\bf r}_i)  \;\;\;\;\langle\texttt{gs}|\hat{S}_i^y|\texttt{gs}\rangle\;=0\;\;
\;\;\langle\texttt{gs}|\hat{S}_i^z|\texttt{gs}\rangle\; = m \cos ({\bf Q}\cdot {\bf r}_i);
$$

\noindent while  the local magnetization $m$ and the condensate of bosons are related  by

\begin{eqnarray}
\langle\phi_c|{b}_{{\bf k}\uparrow}|\phi_c\rangle&=&\left(\frac{N_s m}{2}\right)^{\frac{1}{2}} \;(\delta_{{\bf k},\frac{\bf Q}{2}}+\delta_{{\bf k},-\frac{\bf Q}{2}}) \\
\langle\phi_c|{b}_{{\bf k}\downarrow}|\phi_c\rangle&=& i \left(\frac{N_s m}{2}\right)^{\frac{1}{2}} \; (\delta_{{\bf k},-\frac{\bf Q}{2}}-\delta_{{\bf k},\frac{\bf Q}{2}}) \nonumber,
\end{eqnarray}
\noindent which in real space implies a mean value of the spinors  of the form

$$
 {\langle\phi_c|{b}_{i\uparrow}|\phi_c\rangle\choose\langle\phi_c|{b}_{i\downarrow}|\phi_c\rangle}=\sqrt{2m} \;{\cos \frac{{\bf Q}\;\cdot\; {\bf r}_i}{2}\choose
 \; \sin \frac{{\bf Q}\;\cdot\; {\bf r}_i}{2}}.
$$

\begin{figure}[ht]
\vspace*{0.cm}
\includegraphics*[width=0.45\textwidth,angle=0]{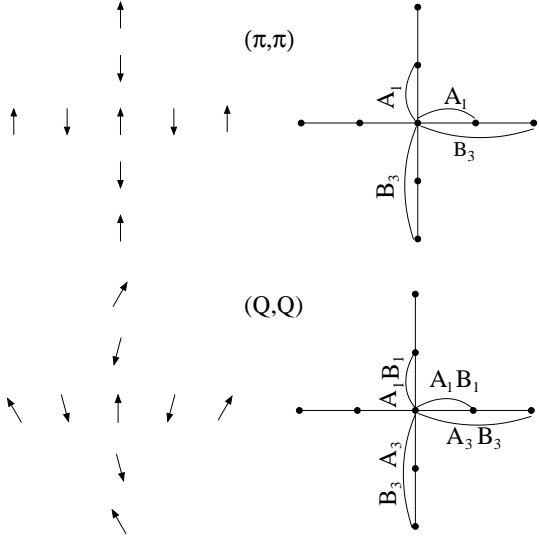}
\caption{Mean field parameter structure corresponding to N\'eel $(\pi,\pi)$ and spiral $(Q,Q)$ correlations for the $J_1-J_3$ model. Only the non vanishing parameters are indicated in each case.}
\label{fig}
\end{figure}

\noindent Replacing these values in (\ref{singlets}) it is obtained the semiclassical expressions for the mean field parameters

\begin{equation}
 A_{\delta}= \langle\phi_c|\hat{A}^{\dagger}_{\delta}|\phi_c\rangle= m \sin \frac{{\bf Q}\cdot \delta}{2}, \;\;\;\;\;\;\;\;\;
B_{\delta}= \langle\phi_c|\hat{B}^{\dagger}_{\delta}|\phi_c\rangle= m \cos \frac{{\bf Q}\cdot \delta}{2}
\label{semiclass}
\end{equation}

\noindent which are consistent with the real nature of the mean field parameters assumed above. This procedure can be performed for a quantum spiral state with magnetization lying in $y-z$ plane. In this case the same semiclassical forms (\ref{semiclass}) are recovered but with $\langle\phi_c|\hat{A}^{\dagger}_{\delta}|\phi_c\rangle$ imaginary pure. It is interesting to note that both mean field solutions are related by a global gauge transformation ${b}_{i\sigma}\rightarrow e^{i \theta} {b}_{i\sigma}$ with $\theta=-\pi/4$. On the other hand, complex values of the mean field parameters $A_{\delta}$ and $B_{\delta}$  can be related to the existence  of non coplanar magnetic or chiral spin liquid states which will be not studied in the present work. For a detailed study of the complex solutions see ref. \cite{messio13}.\\      
\indent Using (\ref{semiclass}), the semiclassical magnetic structures are related to the mean field parameters in the following way (see figure \ref{fig}): a) for N\'eel ${\bf Q}=(\pi,\pi)$ order, $A_{1x}=A_{1y}=A_1\neq0$ and $B_{1x}=B_{1y}=B_1=0$, while $A_{3x}=A_{3y}=A_3=0$ and $B_{3x}=B_{3y}=B_3\neq0$; b) for spiral ${\bf Q}=(Q,Q)$ order, $A_{1x}=A_{1y}=A_1\neq0$, $B_{1x}=B_{1y}=B_1\neq0$, $A_{3x}=A_{3y}=A_3\neq0$ and $B_{3x}=B_{3y}=B_3\neq0$. We have found that this parameter structure is the same for
the LRO and SRO cases, regardless of the quantum or thermal nature of the fluctuations.      
It is worth to stress that for a N\'eel phase frustration $J_3$ is taken into account through the parameter $B_{3}$, whereas for the one operator scheme of decoupling there is no mean field parameter sensitive to frustration since $A_3=0$ (its physical consequence  is clearly reflected in the local magnetization, see figure \ref{fig2}). 

\indent To study $SU(2)$ broken symmetry states, the self consistent equations (12) must be re-calculated taking into account explicitly the condensate    (\ref{condens}) in the thermodynamic limit. The new set of self consistent equations results  

\numparts
\begin{eqnarray}
A_{\delta}\!\!\!\!&=& m \;\sin\frac{\bf Q \cdot\delta}{2}+ \small{\int_{\bf k}} \; \frac{\gamma^A_{\bf k}}{\omega_{\bf k}} \; \sin{\bf k}\cdot {\delta} \;{\it d} {\bf k}  \label{SU2a}\\
B_{\delta}&=& m \; \cos\frac{\bf Q \cdot\delta}{2}+\!\!  \small{\int_{\bf k}}\;\frac{\gamma^B_{\bf k}+\lambda}{\omega_{\bf k}} \;\cos{\bf k}\cdot {\delta} \;{\it d} {\bf k} \label{SU2b}\\
S+\frac{1}{2}&=&m +\!\!  \small{\int_{\bf k}}\;\frac{\gamma^B_{\bf k}+\lambda}{\omega_{\bf k}} \;{\it d} {\bf k}. \label{SU2c}
\end{eqnarray}
\endnumparts
\noindent In addition to the parameters $A_{\delta}$, $B_{\delta}$, and $\lambda$, the magnetization $m$ enters as a new self-consistent parameter. From a comparison  with (12) it follows that the condensate components of  (17)  correspond to the separate treatment of the singular modes ${\bf k}=\pm\frac{\bf Q}{2}$ of the relation dispersion $\omega_{\bf k}$ whereas the sums of (12) are transformed into integrals, as usually presented in the literature \cite{sarker89}. On the other hand, the magnon excitations of the quantum spiral state is obtained by computing the dynamical magnetic structure factor \cite{arovas88,auerbach94}. Here the spectrum of the 
$S^{\dagger}_{\bf k }$  excitations is composed by a pair-spinon continuum with the lowest energy process consisting of  destroying one Schwinger boson $b_{\pm \frac{\bf Q}{2}\downarrow}$ from the condensate and creating another one  $b^{\dagger}_{{\bf k}\pm \frac{\bf Q}{2}\uparrow}$ in the normal fluid part \cite{mezio11}. Given that $\omega_{\pm\frac{\bf Q}{2}}=0$, the energy cost of such a spin-$1$ excitation with momentum ${\bf k}$ is $\omega_{{\bf k}\pm\frac{\bf Q}{2}}$.\\  
The relation dispersion of the spin-$1$ excitation in the large $S$ limit results 
\begin{equation}
 \omega_{{\bf k}\pm\frac{\bf Q}{2}}= {S}\sqrt{ [J_{\bf k}-J_{\bf Q}] [J_{{\bf k}\pm{\bf Q}}-J_{\bf Q}]},
\label{relsemiclass}
\end{equation}

\noindent where (\ref{semiclass}) has been replaced in the shifted spinon dispersion $\omega_{{\bf k}\pm\frac{\bf Q}{2}}$,  $\lambda\!\!\!=\!\!\!-SJ_{\bf Q}$ and $J_{\bf k}= \sum_{\delta} J_{\delta} e^{i {\bf k}.\delta}$. The two possible relation dispersions, $\omega_{{\bf k}+\frac{\bf Q}{2}}$ and $\omega_{{\bf k}-\frac{\bf Q}{2}}$, do not coincide with the semiclassical linear spin wave (LSW) expression

\begin{equation}
 \omega^{LSW}_{{\bf k}}= {S}\sqrt{ [J_{\bf k}-J_{\bf Q}] [(J_{{\bf k}+{\bf Q}}+J_{{\bf k}+{\bf Q}})/2-J_{\bf Q}]}.
\label{relsw}
\end{equation}

\noindent In fact, to recover the conventional spin wave result singlet and triplet mean field parameters must be introduced \cite{chandra90a}. Nonetheless both, (\ref{relsemiclass}) and (\ref{relsw}), have the same zero energy {\it star} modes ${\bf k}=(0,0),(\pm Q,\pm Q),(\pm Q,\mp Q)$ \cite{chandra90a}. For a given spiral order $(Q,Q)$ it is expected only three zero Goldstone modes related to the complete rupture of the $SO(3)$ symmetry; whereas the spurious zero modes  $(\pm Q,\mp Q)$ reflect the lattice symmetry in the spectrum. For example, the spiral $(Q,Q)$ is related to the spirals $(Q,-Q)$ and $(-Q,Q)$ by a global rotation combined with a reflexion about $y$ and $x$, respectively \cite{capriotti04}. In the quantum $S=\frac{1}{2}$ case, however, after the iterative procedure, the SBMF dispersion recovers the correct Goldstone mode structure at ${\bf k}=(0,0),(\pm Q,\pm Q)$ for spiral antiferromagnets; whereas in the spin wave theory the remotion of the spurious zero modes requires to go beyond  the harmonic approximation \cite{rastelli92}. Regarding the functional form of the physical dispersion one could take the minimum of $\{\omega_{{\bf k}+\frac{\bf Q}{2}},\omega_{{\bf k}-\frac{\bf Q}{2}}\}$ as the lowest energy excitation for each ${\bf k}$. Nonetheless, we have recently shown that for the $120^{\circ}$ N\'eel order of the spin-$\frac{1}{2}$ triangular antiferromagnet it is possible to recover the correct relation dispersion --found with series expansions \cite{zheng06} and LSW plus $\frac{1}{S}$ corrections \cite{chernyshev09}-- by a proper reconstruction based on the shifted spinon dispersions parts of $\omega_{{\bf k}\pm\frac{\bf Q}{2}}$  that concentrate the greater spectral weight of the dynamical structure factor \cite{mezio11}. It is worth to stress that at the mean field level the  two spinons building up the magnon-like excitation are free but, after corrections to the SBMF, it is expected low energy tightly bound pairs of spinons  merging  from the continuum \cite{mezio12}.

\section{Results}

\subsection{Zero temperature quantum phase diagram}

\noindent To obtain the zero temperature quantum phase diagram of the $J_1-J_3$ model for arbitrary $S$ we have computed numerically the self consistent equations (17) as follows.
Using (\ref{semiclass}), a classical structure  --$A^{0}_{\delta}$, $B^{0}_{\delta}$. $m^0=S$, and  ${\bf Q}^0$-- is replaced in the spinon relation dispersion (\ref{omega}), in order to get the value of $\lambda^{0}$ that makes the spinon dispersion gapless, $(\gamma^{B^{0}}_{\pm{{\bf Q}^0}/{2}}+\lambda^0)^2 = |\gamma^{A^{0}}_{\pm{{\bf Q}^{0}}/{2}}|^2$. From (\ref{SU2c}) it is obtained $m^0$ and then  $A^{0}_{\delta}, B^{0}_{\delta}, {\bf Q}^0, \lambda^{0}$ and $m^{0}$ are plugged in (\ref{SU2a}) and (\ref{SU2b}) to obtain the new parameters $A^{1}_{\delta}, B^{1}_{\delta}$. Noting that the new minimum  ${\bf k}_{min}$ of $\omega_{\bf k}$ is related to the new spiral pitch as ${\bf Q}^{(1)}=2{\bf k}_{min}$, the iteration is continued until the process converges.              
Depending on the quantum fluctuation strength, which can be measured by the value of $S$, there are solutions with N\'eel and spiral correlations but with $m=0$. We have called these solutions  short range order SRO $(\pi,\pi)$ and SRO $(Q,Q)$, respectively. \\  

\begin{figure}[h]
\vspace*{0.cm}
\includegraphics*[width=0.45\textwidth,angle=-90]{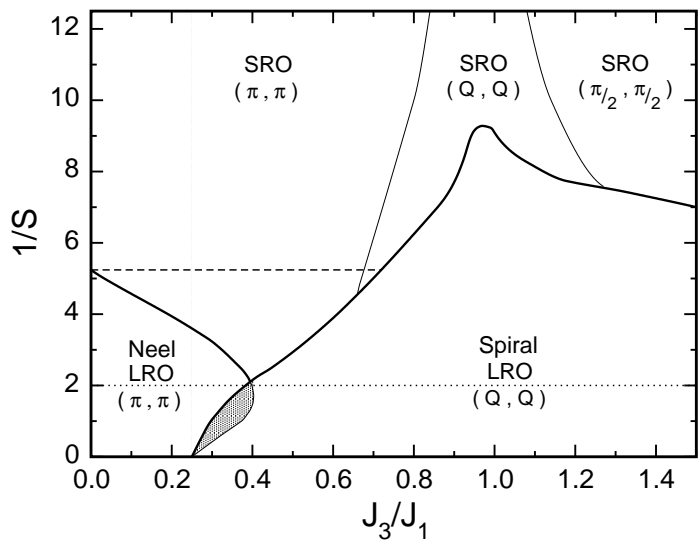}
\caption{Magnetic phase diagram for the $J_1-J_3$ model predicted by the SBMF. Solid lines represent continuous or second order transitions. Thin lines denote disorder lines between different SRO regimes. The hatched area is a metastable N\'eel region and the dotted line indicates the $S=1/2$ case. The dashed horizontal line corresponds to the SBMF prediction within the one singlet decoupling for the N\'eel phase (see text).}
\label{fig1}
\end{figure}

\noindent In figure \ref{fig1} is shown the phase diagram predicted by the SBMF  for all spin and several frustration values:\\

\noindent {\it Long range order regimes}. For $S=\infty$ it is recovered  the classical continuous transitions at $J_3/J_1=0.25$ between LRO N\'eel and LRO spiral phases \cite{locher90}. As $S$ is decreased there is an enhancement of the stability of the N\'eel phase accompanied by a similar reduction of the stability of the spiral phase. This behavior was predicted some time ago using symmetry arguments \cite{ferrer93}. At the transition line of this regime (solid line) the magnetic wave vector change continuously from $(\pi,\pi)$ to incommensurate spiral orders as frustration is increased. 
For spin values $S \gtrsim \frac{1}{2}$ there is a metastable N\'eel region characterized by a reentrance shown in the hatched area of figure \ref{fig1}. This behavior is characteristic of the non trivial interplay between frustration and quantum fluctuations taken into account by the two singlet operator scheme. In particular it has been already found with the same approximation in related models like the $J_1-J_2$  or the $J_2=2J_3$ line of the $J_1-J_2-J_3$ models on the square \cite{flint08,mila91,ceccatto93} and on the honeycomb \cite{mattsson94,cabra11} lattice. If the one singlet operator scheme is applied the solid line delimiting the LRO N\'eel phase should be replaced by the dashed horizontal line of figure \ref{fig1}, missing completely the effect of frustration for the N\'eel phase \cite{read91,chung01}.   The reason of this artefact has already been discussed in Sec. \ref{sec}.
For spin values $S \lesssim0.5$, the continuous transition turns out a second order transition between LRO and SRO states. \\

\noindent {\it Short range order regimes}. The study of the phase diagram for the non physical $S<\frac{1}{2}$ is interesting as one can get an insight of the possible quantum effects beyond the mean field approximation for the physical case ($S=\frac{1}{2}$). In these regimes successive SRO transitions take place across the disorder lines \cite{selke92}(thin lines), $(\pi,\pi) \leftrightarrow (Q,Q)\leftrightarrow (\frac{\pi}{2},\frac{\pi}{2})$, as frustration is varied.  
Here the mean field solutions can be related to the large $N$ limit solutions, $\kappa=\frac{2S}{N}$, where spinons are exactly free only for $\kappa=0$. Inclusion of finite $N$ fluctuations may change drastically the nature of the ground state and the excitations. In this sense, effective gauge field theories  predict that a commensurate SRO ground state is unstable toward a valence bond solid order with confined spinons  while in the incommensurate SRO case a $Z_2$ spin liquid state with deconfined spinons is stabilized \cite{read91}. This physical picture, of course, is beyond the scope of the mean field approximation whose main weakness resides in the relaxation of the local constraint. 
For the regime $S\rightarrow0$  we have found that the two thin lines, separating SRO $(\pi,\pi)$ and $(\frac{\pi}{2},\frac{\pi}{2})$ states, converge into one line (not shown in figure \ref{fig1}) at about $J_3/J_1\sim 1$. For $J_3/J_1 <1$ ($J_3/J_1>1$)  only $A_{1x}=A_{1y}=A_1\neq0$  ($A_{3x}=A_{3y}=A_3\neq0$) survives, respectively. These states that only form singlet bonds $A_{\delta}$ along the links of largest $J_{\delta}$ coincides with a family of solutions coined {\it greedy bosons}, found within the context of the large $N$ theory for $\kappa\rightarrow0$ \cite{tchernyshyov06}. Furthermore, this kind of solutions are in agreement with the upper bounds for the mean filed parameters, $|A_{ij}|\leq {2S}+\frac{1}{2}$ and $|B_{ij}|\leq S $, recently pointed out in \cite{messio13}. On the other hand, it is noticeable the ample room of stability for the SRO $(\pi,\pi)$ phase. In fact, the extended line transition between LRO spirals and  SRO N\'eel $(\pi,\pi)$ phases, about $0.4 \lesssim  J_3/J_1  \lesssim 0.65$, implies a tendency of quantum fluctuations to form commensurate magnetic correlations which in turn will favour valence bond solid states \cite{read91}.
Based on our previous works \cite{trumper97}, we can safely estimate that Gaussian fluctuations  will increase the stability of the SRO $(\pi,\pi)$ and SRO $(Q,Q)$ pushing the LRO $(\pi,\pi)$ and $(Q,Q)$ phases toward higher values of $S$, and thus opening an intermediate disordered window with probably a valence bond solid  or a $Z_2$ spin liquid character.\\
 
\begin{figure}[h]
\vspace*{0.cm}
\includegraphics*[width=0.45\textwidth,angle=-90]{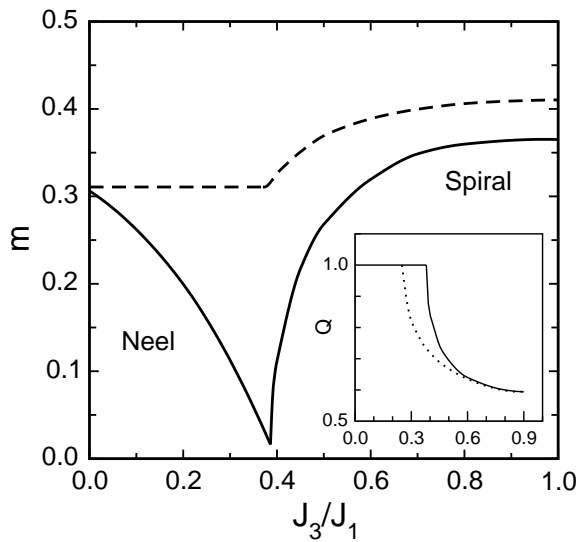}
\caption{Local magnetization $m$ as a function of frustration for the case $S=\frac{1}{2}$. The solid line is for the two singlet operators scheme and the dashed line is for the one singlet operator scheme. In both schemes the transition point occurs at the value $(J_3/J_1)_c\sim0.38$. Inset: $Q$ value, in units of $\pi$, of the magnetic wave vector $(Q,Q)$ versus frustration.}
\label{fig2}
\end{figure}

\noindent{\it Spin $S=\frac{1}{2}$ case}. These results are particularly interesting due to the further comparison with the available numerical studies. In figure \ref{fig2} is plotted the local magnetization versus frustration  for $S=\frac{1}{2}$. There is a continuous transition from N\'eel to spiral phases that turns out quasi-critical at $(J_3/J_1)_c\sim0.38$ with a quite small local magnetization $m\sim 0.015$. In the same figure \ref{fig2} is shown in dashed line the prediction of the one singlet operator $\hat{A}_{ij}$ scheme. Although the transition occurs at the same point, the approximation fails to describe the frustration effects for the N\'eel phase as discussed above for the $\frac{1}{S}$ phase diagram (figure \ref{fig1}). In the inset of figure \ref{fig2} is shown the continuous variation of the magnetic wave vector ${\bf Q}$  with frustration (solid line) where a strong quantum renormalization with respect to the classical value \cite{locher90} (dotted line) is observed. For spiral phases, both scheme of decoupling, one and two singlet operators, predict the same value of ${\bf Q}$ (solid line). 
Regarding the numerical studies for $S=\frac{1}{2}$, they predict the existence of an intermediate disordered regime in the range $0.4 \lesssim J_3/J_1\lesssim 0.8$ with, probably, SRO plaquette and  SRO spiral regimes between LRO N\'eel and LRO spiral phases \cite{leung96,mambrini06,reuther11}; while for the special case $J_3/J_1\simeq 0.5$ there is evidence of an homogeneous spin liquid state \cite{capriotti04a}.  From our previous works \cite{trumper97}, we again estimate that corrections to the mean field will open a disordered window with SRO $(\pi,\pi)$ correlations around the critical value $(J_3/J_1)_c\sim0.38$. By noting that the mean field on site spin fluctuations  $<\hat{{\bf S}}^2_i>=\frac{3}{8} 2S(2S+2)$ do not coincide with the expected value $S(S+1)$, one can choose $S$ in order to adjust the correct local spin fluctuations \cite{messio13}. This procedure gives a spin value $S^*=\frac{1}{2}(\sqrt{3}-1)\sim 0.366$ that, from inspection of figure \ref{fig1} at $\frac{1}{S^*}\sim 2.73$, implies a SRO N\'eel region within the range $0.35\lesssim  J_3/J_1 \lesssim 0.5$. Since these states have a tendency to form valence bond solid states \cite{read91} we conclude that a reasonable agreement with numerical results \cite{reuther11} will be found. However, to recover the homogeneous spin liquid state found at $J_3/J_1=0.5$ one should 
improve the calculation, for example, implementing the local constraint exactly. Recent variational Monte Carlo studies based on SBMF ansatz \cite{sorella12} predict a  $Z_2$ spin liquid state in the disordered regime of the $J_1-J_2$ model, even in the absence of spiral SRO \cite{read91}. Therefore, in agreement with \cite{capriotti04a}, we also expect the probable realization of a $Z_2$ spin liquid in the disordered region of the $J_1-J_3$ model. Recently, similar features have been found using the same approximation for the phase diagram of the  $J_1-J_2$ model on the honeycomb lattice \cite{zhang13}. 
\subsection{Finite temperature phase diagram}

The finite temperature phase diagram is obtained by solving the self consistent equations (12)  with the mean field  parameters $A_{\delta}$, $B_{\delta}$, and $\lambda$. Here, in agreement with the Mermin-Wagner theorem, the magnetization $m$ gives always zero.
This rotational invariant solutions correspond to the renormalized classical regime with an exponential decay of the spin-spin correlation functions \cite{yoshioka91}. In particular, we are mainly interested in the SRO spiral phases since at finite temperature they  break the discrete $Z_2$ symmetry  relating the $(Q,Q)$ and $(Q,-Q)$ phases. In fact, classical Monte Carlo results \cite{capriotti04} predict a $Z_2$ broken symmetry phase that belongs to the Ising universality class characterized by the {\it nematic} order parameter 

\begin{equation}
\sigma=\; \langle \hat{\bf S}_1\cdot\hat{\bf S}_3-\hat{\bf S}_2\cdot\hat{\bf S}_4\rangle,
\label{sigma}
\end{equation}

\noindent where the numbers denotes the sites of a single square plaquette ordered in the cyclic form $(1,2,3,4)$ \cite{capriotti04}. Besides of giving a measure of spiral correlations --it vanishes for N\'eel correlations-- it is easy to see that the order parameter $\sigma$ assumes opposite signs for $(Q,Q)$ and $(Q,-Q)$ correlations. To compute $\sigma$ within the SBMF theory it is enough to resort to (\ref{int}), whence $\sigma$ is written in terms of second neighbours correlations as 

\begin{equation}
\sigma= {B}_{13}^2-{A}_{13}^2 - {B}_{24}^2 + {A}^2_{24}.
\label{orderpar}
\end{equation}

\noindent  Although the mean field parameters are the $A$'s and $B$'s to first and third neighbours, it is 
possible to calculate ${B}_{13}$, ${B}_{24}$, ${A}_{13}$, and ${A}_{24}$ by solving first the self consistent equations (12) and then compute (\ref{selfa}) and (\ref{selfb})  with the vector $\delta$ connecting   second neighbours $(1,1)$ and $(1,-1)$. On the other hand, by plugging in the semiclassical expressions (\ref{semiclass}) the order parameter results 

$$
\sigma=-2S^2 \sin {Q}_x . \sin {Q}_y,
$$

\noindent where the sign difference between $(Q,Q)$ and $(Q,-Q)$ states is evident, as expected. \\

\begin{figure}[ht]
\vspace*{0.cm}
\includegraphics*[width=0.45\textwidth,angle=-90]{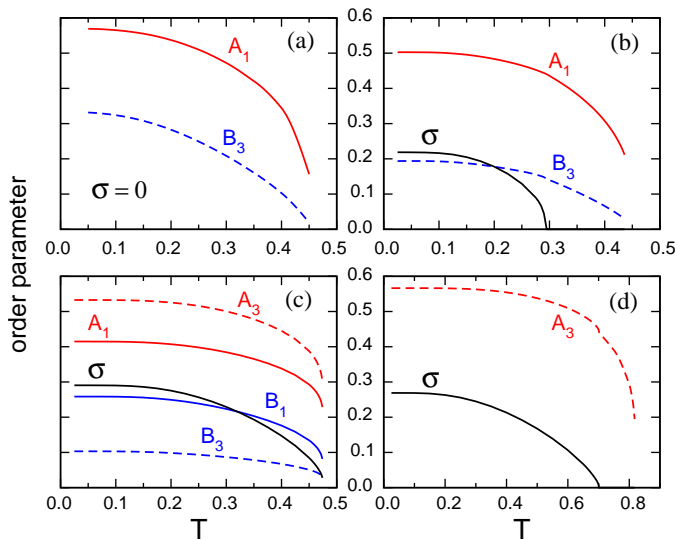}
\caption{Mean field and nematic order parameters versus temperature for several frustration values. (a) $J_3/J_1=0.3$, (b) $J_3/J_1=0.6$, (c) $J_3/J_1=1$ and (d) $J_3/J_1=1.8$ }
\label{correl}
\end{figure}

\begin{figure}[ht]
\vspace*{0.cm}
\includegraphics*[width=0.4\textwidth,angle=0]{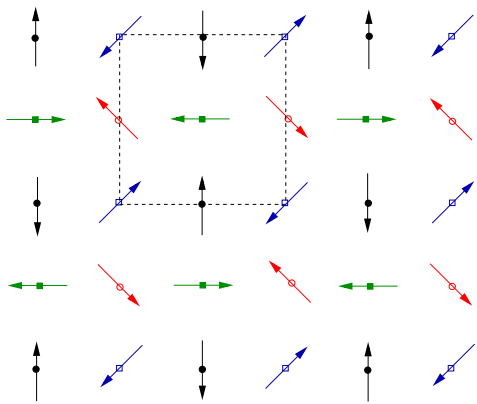}
\caption{Schematic magnetic structure corresponding to the $(\pi,\pi)_4$ phase composed by four decoupled third neighbours sublattices with N\'eel correlations each one.}
\label{pipi4}
\end{figure}

\noindent Depending on the frustration value  we have found different regimes as temperature is increased from the zero temperature ground states. In figure \ref{correl}(a) is shown the temperature dependence of the non zero parameters $A_1$ and $B_3$ corresponding to a N\'eel phase at $J_3/J_1=0.3$.
The parameters decrease monotonously giving rise to a SRO N\'eel phase until $T\sim0.45$. Beyond this temperature the SBMF gives a perfect paramagnet with all the mean field parameters equal to zero. 
Starting from a spiral ground state two different temperature behaviour are observed. On one hand, for $0.38<J_3/J_1<1$, the phase with SRO spiral phase undergoes a transition to SRO N\'eel phase as temperature increases, since fluctuations above a collinear SRO can minimize more efficiently the free energy. This behavior, already observed in related models \cite{hauke10}, is shown in figure \ref{correl}(b) for $J_3/J_1=0.6$. Here the spiral correlations signalled by $\sigma\neq0$ persist until $T\sim0.3$, while for higher temperatures SRO N\'eel correlations are stabilized --$A_1,B_3\neq0$--  until the value $T\sim0.45$ is reached. On the other hand, for $J_3/J_1>1$, before reaching the paramagnetic phase there is again an intermediate collinear phase that we have named $(\pi,\pi)_4$ because it is composed by four decoupled third neighbours sublattices  with SRO N\'eel correlations each one (see figure \ref{pipi4}).  In this way the free energy can be more efficiently minimized since thermal fluctuations above such a decoupled collinear AF SRO between third neigbours optimize both, internal energy and entropy. This is shown in figure \ref{correl}(d) for $J_3/J_1=1.8$ where only the AF  mean field parameter $A_3$ survives along with a weaker ferromagnetic correlations between fifth neighbours $B_5$ (not shown in the figure), and so forth, within the range $0.7< T <0.8$. In figure \ref{correl}(c) is shown the special case $J_3/J_1=1$ where there is a direct transition from a SRO spiral phase  to a perfect paramagnet at around $T=0.45$.\\

 The jumps of $A_1$ and $A_3$ found at this temperature (figures \ref{correl}) are due to the difficulty to solve numerically the constraint equation around $T=0.45$. Actually, on approaching from high temperatures, it can be shown analytically that in certain limits $A_1$ and $A_3$ go continuously to zero \cite{flint08}. In this regime all mean field parameters are zero and the constraint (\ref{selfc}) implies

\begin{equation}
\omega_{\bf k}=\lambda= T\ln(1+\frac{1}{S}). 
\label{hight}
\end{equation}

\noindent Then, assuming that in the limit $J_1>>J_3$ the first mean field parameter that switches on is $A_1$ with its semiclassical form, the equation (\ref{selfa}) yields

\begin{equation}
 \frac{1}{J_1}= \frac{1}{2N} \sum_{\bf k} \frac{(\sin^2 k_x+\sin k_y \sin k_x)} {\omega_{\bf k}} (1+2n_{\bf k}). 
\label{Ahight}
\end{equation}

\noindent Replacing (\ref{hight}) and carrying on the two dimensional integral of (\ref{Ahight}) it is obtained the critical temperature 

$$
T^*_1= \frac{J_1}{2} \frac{(\frac{1}{2}+S)}{\ln (1+\frac{1}{S})}.
$$      

\begin{figure}[h]
\vspace*{0.cm}
\includegraphics*[width=0.45\textwidth,angle=-90]{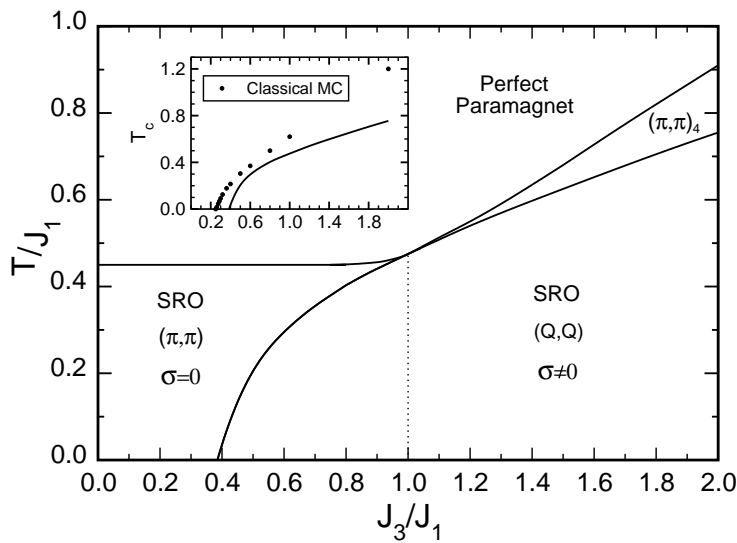}
\caption{Finite temperature phase diagram for the $S=\frac{1}{2}$ case of the $J_1-J_3$ model. Inset: critical temperature $T_c$ for the $Z_2$ broken symmetry phase versus frustration predicted by classical Monte Carlo \cite{capriotti04} (dots) and SBMF (solid line). }
\label{tdiagram}
\end{figure}

\noindent For $S=\frac{1}{2}$ this temperature, $T^*_1\simeq 0.45$, coincides with the horizontal boundary between the paramagnetic and the SRO N\'eel phase  ($J_3/J_1< 1$)
of the finite temperature phase diagram (figure \ref{tdiagram}) found numerically. A similar procedure can be done for $A_3$ in the limit $J_3>>J_1$, giving the critical temperature
$$
 T^*_3\sim \frac{J_3}{2} \frac{(\frac{1}{2}+S)}{\ln (1+\frac{1}{S})}. 
$$
Again, for $S=\frac{1}{2}$, gives a linear behavior $T^*_3\sim0.45 J_3$ that agrees with the boundary between the paramagnetic and the $(\pi,\pi)_4$ regime of the finite temperature phase diagram (figure \ref{tdiagram}). On the other hand, the boundary of the $Z_2$ broken symmetry regime has been numerically identified with the temperature $T_c$ where the nematic order parameter $\sigma$ goes to zero. In the inset of figure \ref{tdiagram} is  shown the qualitative good agreement for the critical temperature $T_c$ of the $Z_2$ broken symmetry phase, as a function of frustration,  predicted by classical Monte Carlo and SBMF theory. Given that the SBMF recovers the classical result in large $S$ limit, the slight shift to the right of $T_c$ with respect to classical MC results can be interpreted as the quantum effect for the $S=\frac{1}{2}$ case. Actually, we expect an even marked shift once correction above the SBMF are computed. 

\section{Concluding remarks }

We have investigated the rupture of the discrete and continuous symmetries in the frustrated  $J_1-J_3$ Heisenberg model  using Schwinger boson mean field theory. We have studied in detail both, the $SU(2)$ broken symmetry which have been explicitly related to the condensate part of the ground state wave function and the $Z_2$ broken symmetry related to the rupture of the discrete degeneracy of the $(Q,Q)$ and $(Q,-Q)$ phases. By comparing with the already existent results, we have shown that the two singlet bond operator scheme of the SBMF give confiable results for the zero temperature quantum phase diagram. In particular, this scheme describes correctly the expected effects of frustration in the collinear phase \cite{ferrer93} that are not captured by the one singlet scheme used in the literature \cite{read91}. 
For $S=\frac{1}{2}$, local spin fluctuations considerations allows us to infer a disordered regime $0.35 \lesssim J_3/J_1 \lesssim 0.5$ that qualitatively agrees with recent numerical results \cite{reuther11}. 
Regarding the finite temperature regime, we have found a $Z_2$ broken symmetry phase characterized by the nematic order parameter $\sigma$ with the rotational invariance restored. The behavior of the critical temperature $T_c$ versus frustration agrees qualitatively  well with classical Monte Carlo results \cite{capriotti04}. \noindent Based on these classical MC results, it has been suggested the possible realization of a $Z_2$ spin liquid with nematic order in the limit $T\rightarrow0$ between the N\'eel and spiral phases \cite{capriotti04}. It should be noticed, however, that in principle there is no connection between the $Z_2$ global symmetry of the Ising-like nematic
order parameter $\sigma$ and the $Z_2$ gauge theory of the spin liquid phase. In the context of the low energy effective field theory  the $Z_2$ gauge symmetry corresponds to  the $Z_2$ gauge invariance of some spinor fields, analog to the Schwinger boson spinors, that results from a particular parametrization of the spiral order \cite{powell11,sachdev08}. In the present microscopic SBMF the nature of the studied quantum and finite temperature solutions are of the same kind --with a finite Ising-like nematic order; consequently  it is important to remark that, if exists, the non trivial properties of the $Z_2$ spin liquid state  will appear, for instance, by solving the hard core local constraint exactly. Nonetheless, at present, its implementation within the variational Monte Carlo shows severe limitations allowing to study system sizes up to $6\times 6$ \cite{sorella12,sorella13, tai11}. Another interesting result is the general tendency of thermal fluctuations to stabilize collinear correlations. In particular, we have found transitions from spiral SRO to collinear N\'eel SRO  before reaching the paramagnetic phase: for $J_3/J_1< 1$ short range  N\'eel $(\pi,\pi)$ correlations are favored while for $J_3/J_1> 1$ there is an intermediate phase $(\pi,\pi)_4$ characterized by four decoupled third neighbours sublattices  with SRO N\'eel correlations each one. Classical Monte Carlo are called for the study of the  $(\pi,\pi)_4$ phase.\\
\indent We have shown that the Schwinger boson mean field theory is a simple and versatile tool that, once adequately implemented, is able to recover the main features  of frustrated Heisenberg models such as static, dynamic and finite temperature properties. It would be interesting to extend the study to doped frustrated antiferromagnets within the context of the $t-J$ model \cite{kane89}  where it is known that spiral fluctuations change drastically the hole spectral functions \cite{trumper04}. Furthermore, the two singlet bond operator scheme used in the present work can be properly extended to the study of anisotropic frustrated models. In particular, for the $XXZ$ model on the triangular lattice we have found \cite{ghioldi13} that the SBMF recovers the dispersion relation predicted by the spin wave plus $1/S$ corrections \cite{chernyshev09}. 

\ack{We thank L Capriotti for sending us his Monte Carlo results. This work was supported by CONICET under grant PIP2009 Nro 1948.}

\end{document}